\documentstyle[12pt,twoside,fleqn,espcrc1]{article}

\newcommand{\AmS}{{\protect\the\textfont2
  A\kern-.1667em\lower.5ex\hbox{M}\kern-.125emS}}

\title{Rotating Nuclei at Extreme Conditions:\\
Cranked Relativistic Mean Field Description}

\author{A.\ V.\ Afanasjev\footnote{Alexander von Humboldt
        fellow. Permanent address: Nuclear Research Center, 
                Latvian Academy of Sciences, LV-2169, 
                Salaspils, Miera str. 31, Latvia} and P.\ Ring
        \address{Physik-Department 
                der Technischen Universit{\"a}t M{\"u}nchen, 
                D-85747 Garching, Germany}}

\begin{document}
\maketitle
\begin{abstract}
The cranked relativistic mean field (CRMF) theory is applied for 
the description of superdeformed (SD) rotational bands observed in 
$^{153}$Ho. The question of the structure of the so-called SD band in
$^{154}$Er is also addressed and a brief overview of applications of CRMF 
theory to the description of rotating nuclei 
at extreme conditions is presented.
\end{abstract}
\input epsf

\section{INTRODUCTION}

CRMF theory \cite{KR,KR.93} represents the 
extension of relativistic mean field (RMF) theory to the rotating frame 
and thus provides a natural framework for the description of rotating
nuclei at high spin. Available experimental data on rotating nuclei 
at extreme conditions of large deformation (superdeformation) and 
fast rotation in different mass regions allow to test the
theoretical models (in our case the CRMF theory) in physical 
situations where pairing correlations are expected to play no or 
only a minor role. This is an especially important point considering 
the fact that in the framework of CRMF theory a consistent theoretical 
description of pairing correlations including fluctuations by number 
projection is still in a stage of development.

  Thus a systematic study of SD bands within CRMF theory has been 
undertaken. Detailed investigations have been performed in the 
$A\sim 140-150$ \cite{KR.93,AKR.96,Hung,ALR.98} and in the 
$A\sim 60$ \cite{A60,Zn60SD,Zn68} mass regions. Experimental 
observables as dynamic moments of inertia $J^{(2)}$, kinematic 
moments of inertia $J^{(1)}$ in the $A\sim 60$ mass region, 
absolute ($Q_0$) and relative ($\Delta Q_0$) charge quadrupole 
moments, effective alignments $i_{eff}$ and the single-particle 
ordering in the SD minimum (derived from the analysis of 
effective alignments) have been confronted with results of 
CRMF calculations without pairing. It was shown that this 
theory provides in general good agreement with available 
experimental data.

All these results give us strong confidence that CRMF theory
can be a powerful tool both for the interpretation of experimental 
data and for the microscopic understanding of the behaviour of 
rotating nuclei at extreme conditions. Considerable disagreement
with experiment has so far only been found in the case of 
the 'SD' band in $^{154}$Er \cite{Er154}. In the present article, 
we report on investigations on the structure of SD bands observed 
recently in $^{153}$Ho \cite{Ho153} with the aim to understand better the 
origin of the discrepancies found in the $^{154}$Er case.

\begin{figure}[t]
\epsfxsize 16.0cm
\epsfbox{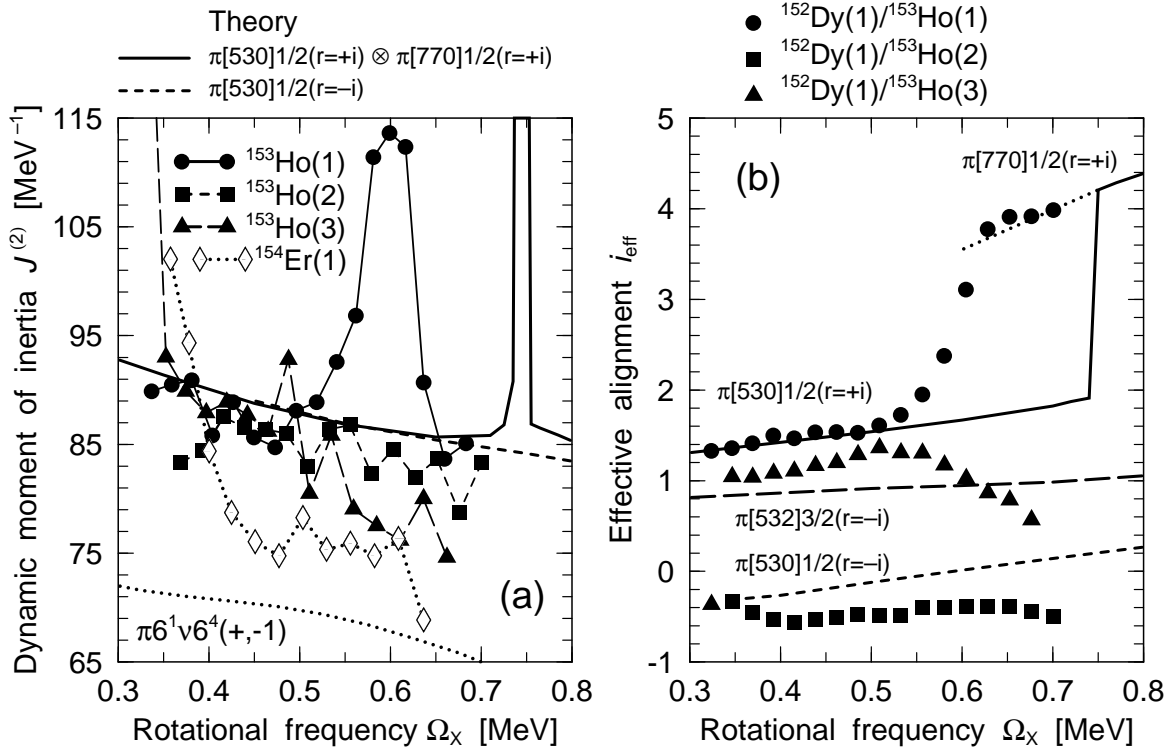}
\vspace{-1.0cm}
\caption{\small (a) Dynamic $J^{(2)}$  moments of inertia
of observed bands (linked symbols) versus the ones 
of calculated configurations. 
Note that $J^{(2)}$($^{153}$Ho(2)) 
is similar to $J^{(2)}$($^{152}$Dy(1)). 
(b) Experimental (symbols) and calculated (lines) effective 
alignments (in units of $\hbar$). 
The effective alignment between bands A and B is defined 
in Ref.\ \protect\cite{Rag91} as
$i_{eff}^{A,B}(\Omega_x) = I_{B}(\Omega_x)-I_{A}(\Omega_x)$.
The band A in the lighter nucleus is taken as a reference,
so the effective alignment measures the effect of the additional
particle(s). The experimental effective alignment between bands
A and B is indicated as ``A/B''. The compared configurations 
differ in the occupation of the orbitals indicated in the 
figure. The dotted line shows the calculated effective alignment 
of the $\pi [770]1/2 (r=+i)$ orbital below the crossing.
The lowest transitions in the observed bands with the transition 
energies of 602.4 keV ($^{152}$Dy(1)), 651.3 keV 
($^{153}$Ho(1)), 713.0 keV ($^{153}$Ho(2)) and 657.0 
keV ($^{153}$Ho(3)) correspond to a spin change 
$26^+ \rightarrow 24^+$, $29.5^- \rightarrow 27.5^-$, 
$30.5^- \rightarrow 28.5^-$ and $28.5 \rightarrow 26.5$,
respectively. Note that in the present formalism only the 
relative spins are "determined" so shifts of all bands in 
steps of $\pm 2\hbar$ could not be excluded, see Ref.\
\protect\cite{ALR.98} for details.}
\label{fig-align}
\vspace{-0.5cm}
\end{figure}

\section{The nuclei $^{153}$Ho and $^{154}$Er.}

{\bf The nucleus $^{153}$Ho.} Three SD bands have been observed in 
$^{153}$Ho \cite{Ho153}. Their structure, as it follows from CRMF 
calculations with the NL1 force \cite{NL1}, is discussed below. 
Considering the large size of the SD shell gaps at $Z=66$ and $N=86$, 
which follows from the doubly magic nature of the $^{152}$Dy SD core 
(conf.\ $\pi 6^4 \nu 7^2$) \cite{AKR.96,TwinDy152},
the occupation of neutron orbitals in the considered configurations 
is kept as in the $^{152}$Dy SD core. Then the configurations based 
on different occupations of the proton orbitals by the 67th proton 
have been calculated. As a result, they are labelled by the proton 
orbital occupied above the $Z=66$ SD shell gap.

{\bf Band 1.}  This band undergoes a band crossing at a
frequency $\Omega_x \sim 0.6$ MeV, where a large increase 
in $J^{(2)}$ is observed (Fig.\ 1a). 
Such a crossing appears also in the lowest SD configuration 
obtained in CRMF calculations (solid line in Fig.\ 1a).
It arises from the crossing between the $\pi [530]1/2(r=+i)$ 
and $\pi [770]1/2(r=+i)$ orbitals which are the lowest proton
orbitals above the $Z=66$ SD shell gap at different 
frequencies (see Fig.\ 4 in Ref. \cite{ALR.98} for single-routhian 
diagrams). For this configuration, the results of our calculations 
are in very good agreement with experiment at low rotational 
frequencies with respect to  $J^{(2)}$ and $i_{eff}$ (see
Fig.\ 1). However, compared with experiment the crossing is 
calculated at somewhat higher frequency and it is sharper. 
The latter feature is possibly due to both the 
deficiencies of the cranking model and the fact that the 
calculations have been carried out as a function of rotational 
frequency but not as a function of spin. The calculated gain 
in alignment at crossing is very close to the measured one and
it would be in perfect agreement with experiment if the 
crossing would have been calculated at the experimental crossing
frequency. The same interpretation of this band has been
obtained also in cranked Woods-Saxon calculations at fixed 
deformation \cite{Ho153}.

{\bf Band 2.} According to the CRMF calculations, we can 
assign to this band the configuration $\pi [530]1/2 (r=-i)$.
Assuming this assignment, the experimental
values of $J^{(2)}$ and $i_{eff}$ are reasonably well
reproduced (see Fig.\ 1). This assignment corresponds
to the one discussed in Ref. \cite{Ho153}. Additional 
confirmation of the interpretation of bands 1 and 2 
could be obtained by a precise measurement of the
charge quadrupole moments $Q_0$ relative to the ones of 
the $^{152}$Dy(1) band. According to the calculations, 
the occupation of the $\pi [530]1/2(r=+i)$,
$\pi [530]1/2 (r=-i)$ and $\pi [770]1/2(r=+i)$
orbitals leads to an increase of $Q_0$ by 0.60 $e$b,
by 0.65 $e$b (both values are calculated at $\Omega_x=0.5$ MeV)
and by 1.15 $e$b (calculated at $\Omega_x=0.8$ MeV), 
respectively.

{\bf Band 3.} The features of this band are difficult 
to explain assuming that the changes of the physical 
observables with respect to the ones of the $^{152}$Dy(1) 
band should be governed by an additional proton. At 
high rotational frequencies, the $J^{(2)}$ moment of 
inertia drops considerably below that of the $^{152}$Dy(1) 
band. This drop is accompanied by the loss in effective 
alignment $i_{eff}$ of $\approx 0.8\hbar$ in the $\Omega_x=0.51-0.68$ 
MeV range (Fig. 1). It was suggested in 
Ref.\ \cite{Ho153} that the occupation of the 
$\pi [523]7/2(r=-i)$ orbital by 67th proton could lead to 
such features. It seems that this interpretation can be 
ruled out since the calculated loss of alignment of 
$\approx 0.2\hbar$ arises from the interaction between 
the $(r=-i)$ signatures of the 
$\pi [523]7/2$ and the $\pi [530]1/2$ orbitals. However, 
the configuration with the $\pi [530]1/2 (r=-i)$ 
orbital occupied is assigned to band 2 which does not 
show an increase neither in $J^{(2)}$ nor in $i_{eff}$ 
expected from such an interaction. A consistent 
interpretation of this band within a pure single-particle 
picture is not found in the CRMF calculations either. 
For example, the effective 
alignment of the $\pi [532]5/2(r=-i)$ orbital 
located above the $Z=66$ SD shell gap (see Fig. 4
in Ref.\ \cite{ALR.98}) is shown in Fig.\ 1b. The 
calculated $J^{(2)}$ moment of inertia of this 
configuration is very close to the one of the
configuration assigned to band 2 (Fig. 1a). 
Although the results of calculations are 
reasonably close to experiment at low frequencies, 
the loss of $i_{eff}$ and the drop in $J^{(2)}$ 
at higher frequencies are not reproduced.

 The occupation of positive parity $\pi [413]5/2$, 
$\pi [404]9/2$ and $\pi [411]3/2$ orbitals located 
above the $Z=66$ SD shell gap (Fig.\ 4 in 
Ref.\ \cite{ALR.98}) has also been considered.
The strongest argument against the interpretation
of the observed bands as based on these orbitals comes 
from the fact that these orbitals have a small signature 
splitting. Thus 
signature partner bands with small 
signature splitting should be observed if 
these orbitals are occupied. 

{\bf The nucleus $^{154}$Er.}  One band has been observed in 
$^{154}$Er and it has been discussed as SD \cite{Er154}. 
Two specific 
features of this band are (i) the  $J^{(2)}$ moment of 
inertia at high frequencies is much lower than the one 
of the $^{152}$Dy(1) band (Fig. 1a), (ii) the effective 
alignment in the $^{152}$Dy(1)/$^{154}$Er(1) pair drops by 
$\approx 2.1\hbar$  in the frequency range $\Omega_x=0.37-0.65$ 
MeV.  These features strongly suggest that this band has a 
smaller number of high-$N$ orbitals occupied  (and thus is 
less deformed) than the $^{152}$Dy(1) band. Considering 
available single-particle orbitals above the $Z=66$
SD shell gap and their impact on physical observables
(as deduced from the analysis of $^{153}$Ho),
it is clear that this band cannot be described as
a 'doubly magic $^{152}$Dy core + 2 additional
protons' system. Indeed, the results of calculations 
for $J^{(2)}$ \cite{AKR.96} and $i_{eff}$ of the  lowest 
SD configurations in this nucleus disagree considerably 
with experiment. The possibility that the observed band 
belongs to a highly-deformed triaxial minimum predicted 
in Ref.\ \cite{Er154WS} has also been checked. Such a minimum 
with $Q_0 \sim 10$ $e$b and $\gamma \sim 9^{\circ}$
exists in CRMF calculations too and it is lower in energy than 
the SD minimum at $I< 60\hbar$. Fig. 1a shows the $J^{(2)}$ moment 
of inertia of one of the configurations ($\pi 6^1 \nu 6^4(+,-1)$)
calculated in this minimum. Although there still is 
disagreement with experiment, the discrepancy is somewhat smaller 
than in the case of the SD configurations. However, it is
difficult to present a specific configuration assignment for 
the observed band. The measurements of the transition quadrupole 
moment of this band will help to resolve the existing problem.

\section{Conclusions}

CRMF theory has been applied for the study of SD bands 
observed in $^{153}$Ho. Bands 1 and 2 are reasonably well 
described, while it was difficult to get a consistent 
interpretation for band 3 in a pure single-particle picture. 
Based on these results it was concluded that the band 
observed in $^{154}$Er and previosly discussed as SD 
is very likely less deformed than the $^{152}$Dy(1) 
band.

   A.V.A. acknowledges support from the Alexander von
Humboldt Foundation. This work is also supported in part
by the Bundesministerium f{\"u}r Bildung und Forschung
under the project 06 TM 875.

\end{document}